# Perancangan UI/UX Aplikasi Sistem Informasi Layanan Administrasi dalam Perspektif Psikologi Menggunakan Metode Prototype

Sania Febriani[1)*)], Tata Sutabri[2)], Megawaty[3)], Leon A. Abdillah[4)]

[1)2)3)4)] Fakultas Sains Teknologi, Universitas Bina Darma
[*)]Correspondence Author: saniafb51@gmail.com, Palembang, Indonesia
DOI: https://doi.org/10.37012/jtik.v9i2.1714

## ABSTRAK

Pelayanan administrasi mahasiswa Universitas Bina Darma masih dilakukan secara konvensional. Mahasiswa menemui dosen untuk meminta dosen membubuhkan tanda tangan dokumen administrasi mereka. Meski demikian, kasus pemalsuan tanda tangan tetap terjadi di Universitas Bina Darma. Masalah ini dapat menyebabkan kerugian materi dan termasuk kategori tindak pidana. Tujuan dari penelitian ini membuat desain *interface* Sistem Informasi Layanan Administrasi (SILASTRI) dengan menerapkan teori psikologi warna, prinsip Gestalt dengan *user experience* yang baik. SILASTRI dirancang untuk mendukung layanan administrasi mahasiswa Universitas Bina Darma. Pengumpulan data melalui observasi, penyebaran kuesioner dan studi kepustakaan. Penelitian ini menggunakan metode *prototype* yang terdiri dari *communication, quick plan, modelling quick design, construction of prototype* dan *deployment delivery & feedback*. Metode *prototype* membuktikan kelayakan teknis dan memvalidasi kegunaan tampilan *user interface* dengan melakukan perkiraan dari suatu *software* sehingga jika terdapat kekurangan bisa segera diperbaiki. Berdasarkan hasil *usability testing* menggunakan Maze yang telah di uji coba oleh 70 orang responden diperoleh nilai maze *usability s*ebesar 89 serta nilai perhitungan SUS sebesar 88 yang masuk kategori *good*. Oleh karena itu dapat disimpulkan rancangan UI/UX aplikasi SILASTRI dengan menerapkan perspektif psikologi memiliki *interface* dan *user experience* yang diterima dengan baik oleh pengguna. Hasil pengujian dan evaluasi ini membuktikan bahwa desain tampilan SILASTRI sudah siap dikembangkan menjadi aplikasi.

**Kata kunci:** Psikologi, UI/UX, Prototype, Administrasi

## Abstract

*Bina Darma University student administration services are still carried out conventionally. Students meet the lecturer to ask the lecturer to sign their administrative documents. However, cases of forged signatures still occur at Bina Darma University. This problem can cause material loss and is included in the category of criminal offense. The aim of this research is to design an Administrative Services Information System (SILASTRI) interface by applying color psychology theory, Gestalt principles with a good user experience. SILASTRI is designed to support student administration services at Bina Darma University. Data collection through observation, distributing questionnaires and literature study. This research uses a prototype method which consists of communication, quick plan, modeling quick design, construction of prototype and deployment delivery & feedback. The prototype method proves technical feasibility and validates the usability of the user interface display by estimating the software so that if there are deficiencies they can be corrected immediately. Based on the results of usability testing using Maze, which was tested by 70 respondents, the Maze usability value was 89 and the SUS calculation value was 88, which is in the good category. Therefore, it can be concluded that the UI/UX design of the SILASTRI application by applying a psychological perspective has an interface and user experience that is well received by users. The results of this testing and evaluation prove that the SILASTRI display design is ready to be developed into an application.*

# PENDAHULUAN

Mahasiswa Universitas Bina Darma dalam melaksanakan kegiatan akademik membutuhkan kelengkapan administrasi sebagai bukti legalitas yang harus dibubuhi tanda tangan Ketua Program Studi dan Dekan Fakultas. Proses pembubuhan dokumen administrasi mahasiswa dengan tanda tangan basah mengharuskan pertemuan secara langsung antara dosen dan mahasiswa. Meski demikian, masih terjadi kasus pelanggaran kode etik atau kecurangan akademik berupa manipulasi tanda tangan dosen yang dilakukan oleh mahasiswa. Kasus pemalsuan tanda tangan dosen ternyata terjadi di beberapa Universitas di Indonesia, salah satunya terjadi di Universitas Bina Darma. Beberapa mahasiswa melakukan pemalsuan tanda tangan dosen pada berkas administrasi akademik. Meskipun perbuatan ini dapat memberi kerugian secara material dan masuk kategori pelanggaran tata tertib namun di Universitas Bina Darma namun belum ada aplikasi yang dapat mengatasi permasalahan tersebut. Masalah lainnya ialah proses layanan administrasi mahasiswa masih dilakukan secara konvensional sehingga mahasiswa tidak dapat mengetahui kapan surat mereka selesai diproses. Akibatnya mereka akan datang secara berkala menemui dosen yang bersangkutan. Hal ini tentunya akan menjadi kendala ketika Ketua Program Studi tidak sedang berada di tempat, serta perlunya jeda waktu dalam menyelesaikan dokumen tertentu berdasarkan tingkat kepentingan atau jenis permasalahan.

Teknologi Informasi dan Komunikasi (TIK) terbukti menjadi ikon kemajuan di abad ke-21 (Abdillah & Kurniasti, 2022). TIK telah memantapkan dirinya sebagai pusat berbagai aktivitas manusia (Abdillah et al., 2023). Interaksi Manusia dan Komputer (IMK) dikatakan sebagai disiplin ilmu yang berhubungan dengan perancangan, evaluasi, dan implementasi sistem komputer interaktif untuk digunakan oleh manusia dan studi tentang fenomena di sekitarnya. Untuk menghubungkan sistem dan manusia diperlukan penghubung sistem atau *interface* (Sutabri, 2012). Interaksi manusia dan komputer (IMK) bertujuan agar sistem *user friendly* dengan nilai *usability* yang tinggi. *Usability* adalah bagian dari elemen *user experience* yang memastikan pengguna dapat menggunakan produk dengan mudah. Untuk menciptakan *user experience* sebuah produk harus disesuaikan antara fitur dengan kebutuhan pengguna hingga pengguna dapat merasa senang dan puas ketika melihat atau menggunakan produk. Psikologi dalam penerapannya berkaitan dengan Interaksi Manusia dan Komputer (IMK) untuk memperkecil terjadi kesalahan manusia ketika berhubungan dengan mesin (*human eror)*. Salah satu teori psikologi yang banyak diterapkan pada perancangan desain antarmuka adalah prinsip Gestalt. Prinsip Gestalt juga dikenal sebagai invarian, yaitu prinsip kesamaan atau kemiripan dalam segi hal bentuk, warna dan ukuran (Lengkong et al., 2021). Prinsip Gestalt secara keseluruhan berkaitan dengan interaksi manusia dan komputer (MacNamara, 2017). Desain yang baik harus menjadi solusi permasalahan, mudah digunakan, dan memenuhi kebutuhan pengguna (Shirvanadi, 2021).

Beberapa contoh dari penelitian sebelumnya yang serupa dengan topik penelitian ini yang pertama dilakukan oleh (Puspa, 2020). Tujuan dari penelitian tersebut adalah membuat *company profile* berupa aplikasi berbasis android sebagai sarana informasi agar calon klien dapat mengetahui tentang perusahaan. Pada perancangan desain digunakan pendekatan teori Gestalt mengenai: *similarity, continuity, closure proximity dan figure.* Penelitian ini menghasilkan tampilan aplikasi mobile yang siap diintegrasikan ke pengembang. Penelitian kedua yang serupa juga menerapkan prinsip Gestalt dilakukan oleh (Kelsun & Kristanto, 2021). Hasil penelitian ini berupa desain aplikasi *mobile* bernama Family Pass. Dari penerapan prinsip Gestalt memvalidasi bahwa seperti aspek pemilihan warna, *font*, dan grid atau *layout* bedampak terhadap respons dan *user experience* ketika mengoperasikan sebuah aplikasi. Hasil akhir dari perancangan ini berupa desain interaktif atau prototipe desain. Ketiga penelitian yang dilakukan oleh (Julian et al., 2023) menghasilkan desain UI/UX aplikasi *mobile* forum diskusi mahasiswa. Dengan menerapan prinsip *proximity dan similarity* membuktikan sebanyak 60% responden mengatakan aplikasi mudah digunakan dari pengujian *usability.*

Selain karena platform akademik Universitas Bina Darma seperti *e-learning,* SisFo, dan informasi profil kampus berbasis *website,* dikatakan teknologi *website* yang responsif dapat diakses dari manapun dan kapapun sehingga pengguna lebih mudah membaca informasi dalam website tersebut (Sutabri et al., 2022). Dari permasalahan tersebut peneliti bertujuan menerapkan prinsip Gestalt untuk merancang desain *interface* aplikasi Sistem Informasi Layanan Administrasi (SILASTRI) berbasis *website* dengan *user experience* yang baik.

## METODE

Pengumpulan data penelitian ini dilakukan dengan tiga teknik yaitu:

1. Observasi

   Peneliti melakukan observasi melalui kegiatan magang selama enam bulan di Program Studi Sistem Informasi Universitas Bina Darma untuk mempelajari dan terlibat langsung dalam pelayanan administrasi mahasiswa.

2. Studi pustaka

   Penulis melakukan kajian studi pustaka dari penelitian terdahulu agar menjadi landasan pengetahuan dengan mengumpulkan dan merumuskan dari berbagai sumber seperti artikel jurnal dan *e-book*. Adapun teori yang dikaji mengenai perancangan, psikologi warna, prinsip Gestalt, *user interface* (UI), *user experience* (UX), layanan administrasi, *website, system usability scale* (SUS), *prototype* dan *unified modelling language* (UML).

3. Kuesioner

   Hasil yang didapat dari kuesioner adalah data deskriptif berupa uraian, tabel atau gambar yang akurat dan kredibel. Data yang diperoleh adalah data sekunder berupa tabel, dan dokumen yang relevan. Dalam penelitian ini yang menjadi populasi adalah mahasiswa aktif Fakultas Sains Teknologi Universitas Bina Darma Palembang. Teknik pengambilan sampel yang digunakan adalah *purposive sampling*. *Purposive sampling* merupakan Teknik penentuan sampel dengan pertimbangan tertentu (Zulkifl et al., 2021). Kriteria-kriteria yang peneliti tetapkan ketika pengambilan sampel adalah: Pertama, mahasiswa S1 semester 6 dan 8 dikarenakan saat penelitian ini dilakukan tengah masuk tahun ajaran genap, selain itu pada tingkat tersebut mahasiswa mulai aktif untuk mengikuti kegiatan akademik, dan mengerjakan tugas akhir, sehingga mereka harus melengkapi administrasi akademik. Kedua, mahasiswa berasal dari Fakultas Sains Teknologi, karena jumlah mahasiswa tersebut paling banyak jika dibandingkan fakultas lain yang ada di Universitas Bina Darma sehingga dapat mewakili populasi. Ketiga, penentuan jumlah responden dipertimbangkan dari jumlah responden *usability testing* yaitu minimal 20 orang (Yuliyana et al., 2019). Menurut Singarimbun dan Effendi (1995) sebanyak 30 orang responden dikatakan cukup untuk mewakili populasi dan menjaga kualitas penelitian (Veni Manik et al., 2021). Selain itu tidak ada aturan baku dalam menentukan responden SUS sehingga banyaknya jumlah responden dapat disesuaikan dengan kebutuhan peneliti (Ependi et al., 2019). Dari pengujian yang telah dilakukan jumlah responden pengujian *usability* sebanyak 70 orang.

Menurut Roger. S. Pressman, Ph. D (2005) *prototype* merupakan pendekatan terbaik untuk mendefinisikan serangkaian tujuan umum untuk perangkat lunak, tetapi tidak mengidentifikasi *input,* pemrosesan, atau *output* yang terperinci. Berikut ini merupakan tahapan yang dilakukan dalam metode prototype:

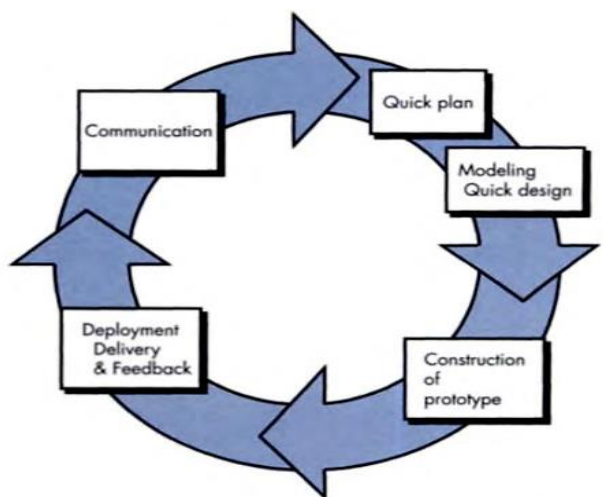

**Gambar 1.** Paradigma *Prototype* Menurut (Pressman, 2005)

1. *Communication*

   Pada tahap c*ommunication*, peneliti menyebarkan kuesioner kepada mahasiswa melalui *google form* sebagai upaya untuk memahami kebutuhan pengguna, serta menerima saran dari para ahli yaitu dosen pembimbing dan dosen penguji mengenai spesifikasi sistem seperti apa yang butuhkan oleh pengguna. Hasil survei dibuktikan bahwa proses sebuah dokumen membutuhkan lebih dari 3 kali pertemuan karena mahasiswa tidak mengetahui kapan tepatnya dokumen tersebut selesai diproses sehingga 23 orang menjawab menemui dosen bersangkutan diesok harinya. Selain itu, diketahui sebanyak 29 orang responden menyatakan aplikasi SILASTRI dapat membantu mereka dan 24 responden memilih tampilan desain yang minimalis.

2. *Quick Plan*

   Dalam tahap *quick plan* merupakan lanjutan dari proses sebelumnya, dengan membuat perencanaan dan pemodelan untuk merepresentasikan entitas yang ditampilkan dan dilihat oleh pengguna. Representasi tersebut berupa diagram *unified modelling language* sistem yang akan menjadi dasar pembuatan *prototype.*

3. *Modelling Quick Design*

   Pada tahap ini dilakukan perancangan *wireframe*. *Wireframe* adalah sebuah kerangka untuk menata suatu item di laman *website* atau aplikasi sebelum proses mendesain tampilan antarmuka. Pada tahap ini diterapkan teori psikologi penggunaan warna dan prinsip Gestalt untuk menata teks, gambar, *layout*, bentuk tombol, navigasi dan sebagainya.

4. *Construction of Prototype*

   Pada tahap *contruction of prototype*, *wireframe* akan dikembangkan menjadi desain *high fidelity* dan *prototyping* interaktif menggunakan *tools* Figma berdasarkan kebutuhan sistem yang telah didefinisikan sebelumnya. Desain *high fidelity* berbeda dari *wireframe* karena desain telah memiliki warna, elemen dan visual dengan tingkat presisi yang tinggi dan menjadi produk final sehingga pengguna bisa berinteraksi dengan *prototype* tersebut.

5. *Deployment delivery & feedback*

   Setelah *prototype* jadi akan diberikan ke pengguna yaitu mahasiswa untuk dilakukan *usability testing* guna mengetahui apakah pengguna dapat menggunakannya dengan mudah atau tidak, serta apakah mengguna memiliki *experience* yang baik saat menggunakan *prototype* SILASTRI. Dengan pembuatan prototipe selain untuk membuktikan kelayakan teknis dan memvalidasi kegunaan antarmuka pengguna, dapat dilakukan perkiraan tampilan dari suatu *software* sehingga apabila ditemukan kekurangan dapat segera diperbaiki (Abdillah et al., 2019). Pengujian dilakukan menggunakan tools Maze yang telah terintegrasi dengan Figma. Kemudian prototype akan dinilai oleh pengguna dengan mengisi kuesioner *system usability scale* (SUS) melalui *google form*. *Feedback* berupa skor *usability testing* dan skala hasil evaluasi SUS terhadap prototype yang telah pengguna coba. *Feedback* responden dapat dijadikan pertimbangan untuk pengembangan *software* lebih lanjut.

## HASIL DAN PEMBAHASAN

Terdapat dua aktor pada *use case diagram* SILASTRI, yaitu mahasiswa dan admin. Pada diagram ini menampilkan fitur apa saja yang dapat digunakan aktor pada aplikasi SILASTRI serta proses yang dilalui. Setelah berhasil *login,* mahasiswa mengklik menu pengajuan pada beranda kemudian sistem menampilkan *form* pengajuan.

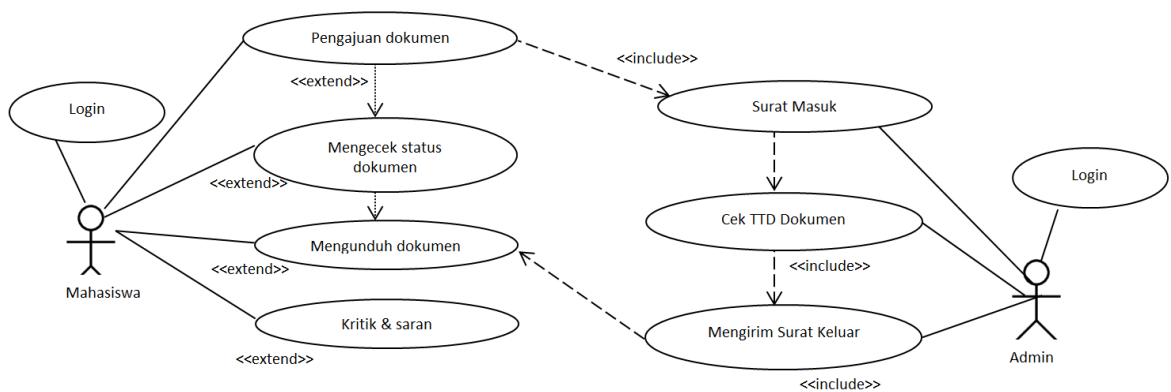

**Gambar 2.** Use Case Diagram

Setelah semua data di isi maka akan terkirim ke surat masuk admin. Admin mengunduh dokumen dan menggunakan fitur cek tanda tangan SILASTRI untuk memverifikasi tanda tangan dosen yang tertera pada dokumen yang diajukan mahasiswa. Setelah diketahui hasilnya adalah asli maka dokumen akan diteruskan kepada Ketua Program Studi atau Dekan Fakultas untuk di bubuhi tanda tangan. Setelah itu admin akan *scan* dokumen tersebut untuk dikirim kembali ke mahasiswa melalui fitur surat keluar. Setelah mengirim pengajuan mahasiswa dapat memantau dokumen mereka melalui fitur cek dokumen, jika sudah selesai maka dokumen dapat di unduh dan menjadi arsip yang dapat diakses kapan pun.

Dalam merancang desain tampilan antarmuka SILASTRI digunakan warna biru sebagai warna dominan. Warna biru memberikan kesan positif yang menyenangkan dan melambangkan keyakinan (Yogananti, 2015). Pendapat ini juga di dukung oleh (Lengkong et al., 2021) yang mengatakan bahwa warna biru memiliki karakteristik kebutuhan akan keteraturan dan arah kehidupan. Penerapa prinsip Gestalt biasa digunakan para pelaku seniman untuk memvalidasi struktur komposisi sebuah pola visual dengan dasar psikologis yang dijadikan sebagai acuan, pendapat ini juga didukung (Puspa, 2020) yang mengatakan Teori Gestalt sering dipakai dalam proses desain dan cabang seni rupa lainnya. Karena itulah pemberian warna dan beberapa elemen seperti logo aplikasi, teks, icon, tombol, dan gambar mengikuti prinsip Gestalt seperti berikut ini:

1. *Law of Symmetry*

    Posisi komponen yang diletakkan secara simetris membuat tampilan menjadi seimbang sehingga konten tidak terlihat menumpuk atau penuh pada suatu sisi saja. Pengguna mendapat pengalaman yang nyaman ketika melihat tampilan desain antarmuka SILASTRI yang dibuat simetris.

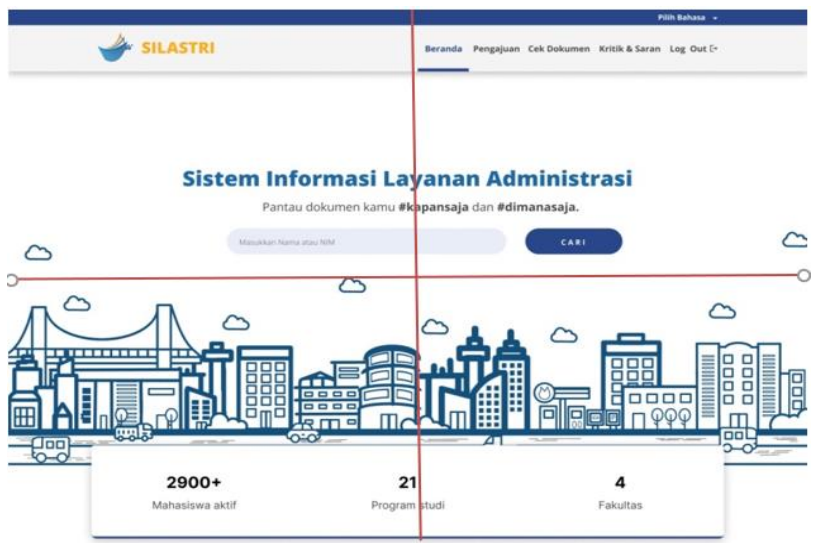

**Gambar 3.** Penerapan Prinsip *Law of Symmetry*

2. *Law of Continuity*

    Penggunaan prinsip ini diterapkan untuk memandu arah pandangan pengguna untuk mengikuti garis lurus yang selaras dari suatu kelompok yang desainer inginkan. Pada halaman dibawah ini diterapkan prinsip *continuity* untuk mempermudah pengguna memperoleh informasi atau menggunakan navigasi dari SILASTRI.

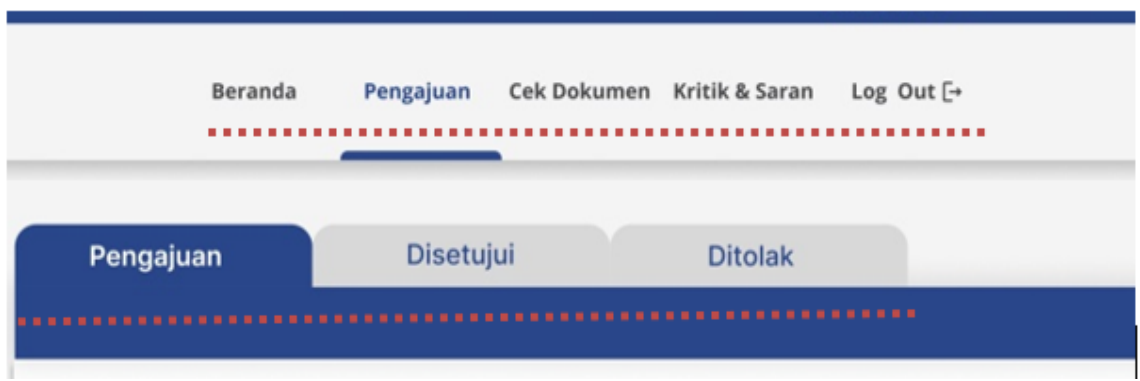

**Gambar 4.** Penerapan Prinsip *Law of Continuity*

3. *Law of Figure Ground*

Pesan *pop-up* merupakan objek perhatian (figur) yang tampil untuk mengarahkan perhatian pertama pengguna, kemudian pengguna melihat *background*. Sehingga pesan dari sistem tersampaikan dengan baik kepada pengguna.

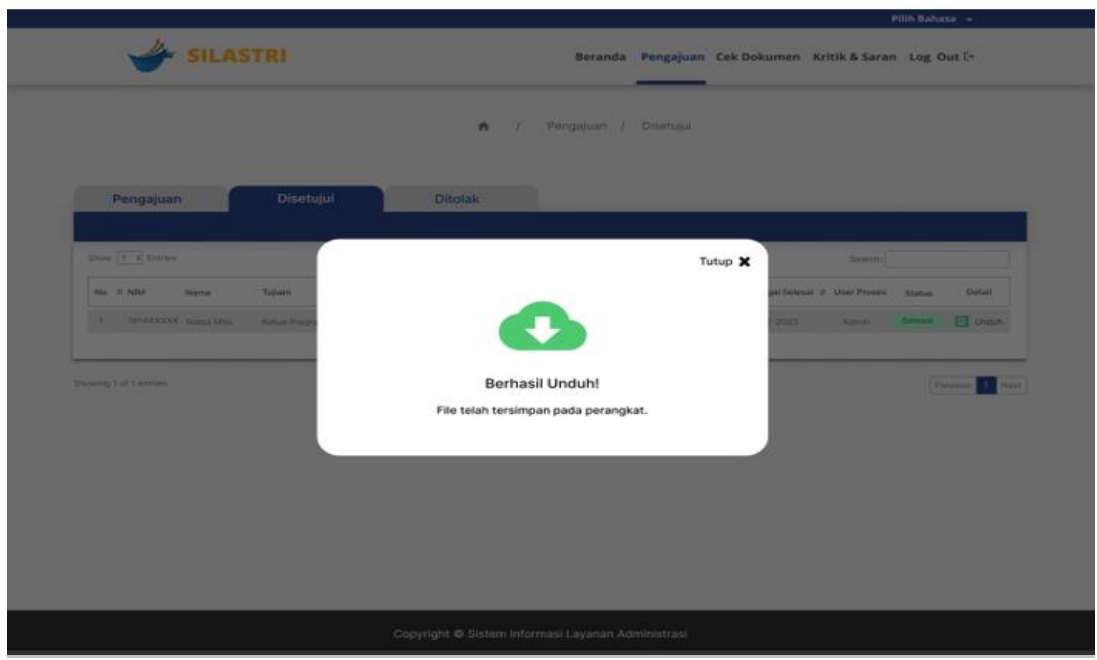

**Gambar 5.** Penerapan Prinsip *Law of Figure Ground*

4. *Law of Common Fate*

Penerapan prinsip ini untuk menciptakan efek pengguna melihat elemen yang bergerak bersama sebagai satu kesatuan. Pada tampilan hasil pencarian dibuat secara vertikal, sehingga pengguna menggulirkan kursor maka tampilan bergerak bersamaan ke atas atau ke bawah.

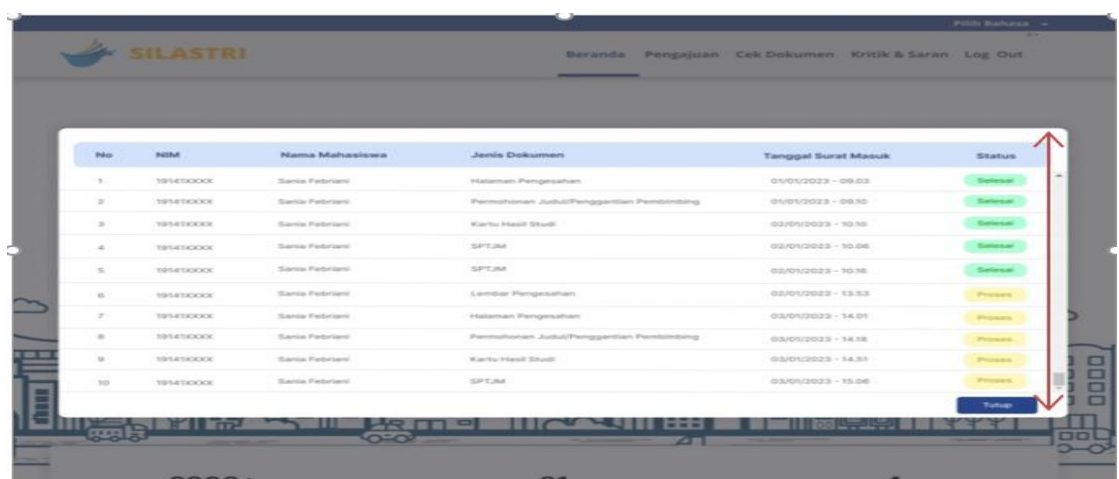

**Gambar 6.** Penerapan Prinsip *Law of Common Fate*

5. *Law of Proximity*

   Penerapan prinsip *proximity* digunakan untuk mengatur kedekatan jarak atau spasi antar elemen agar membuat pengguna mudah mengenali bahwa setiap kelompok mempunyai peranan yang berbeda-beda. Tampilan halaman cek tanda tangan ini terbagi menjadi empat kelompok yaitu *top bar*, identifikasi tanda tangan, navigasi yang terdiri dari bagian menu dan akun.

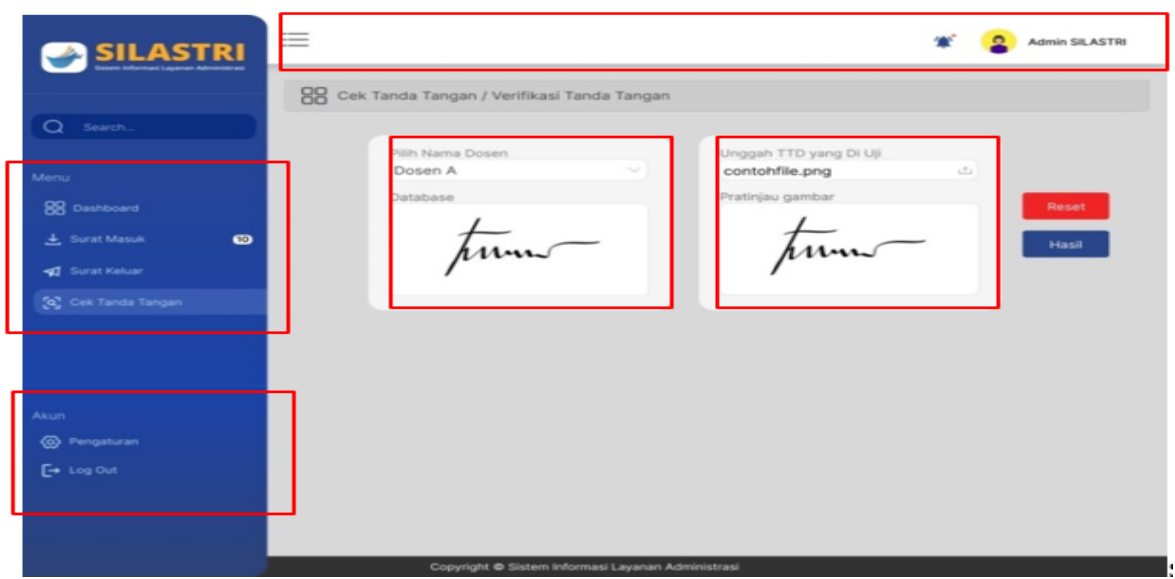

**Gambar 7.** Penerapan Prinsip *Law of Proximity*

6. *Law of Closure*

   Pikiran manusia memiliki memiliki kemampuan untuk melihat bagian yang tidak utuh namun tetap dapat mengenalinya dari suatu desain yang ditampilkan. Karena itu prinsip ini diterapkan pada bagian *input* agar pengguna mengisi dan melengkapi bagian kosong tersebut. Contohnya adalah pada kolom login, input data surat masuk dan input surat keluar. Hal tersebut terlihat pada gambar 8.

7. *Law of Simplicity*

   Suatu desain yang sederhana akan menghasilkan visual yang mudah dipahami bagi yang melihat. Desain tampilan *dashboard* admin dibuat dengan sederhana mungkin agar mudah dipahami pengguna. Hal tersebut terlihat pada gambar 9.

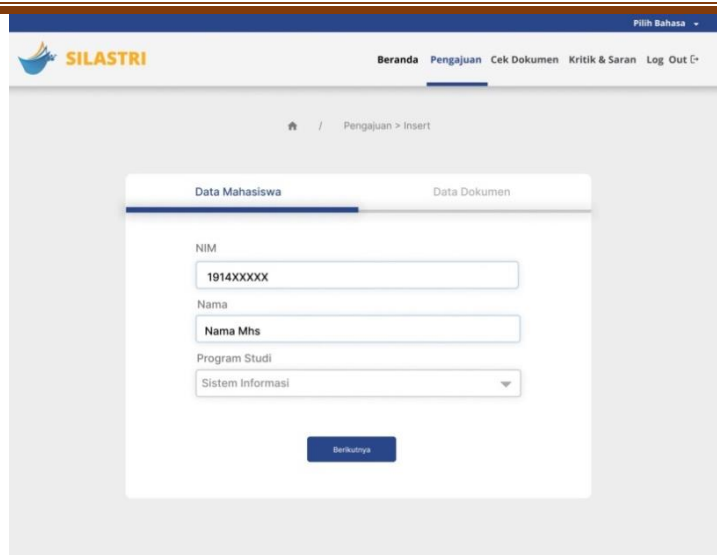
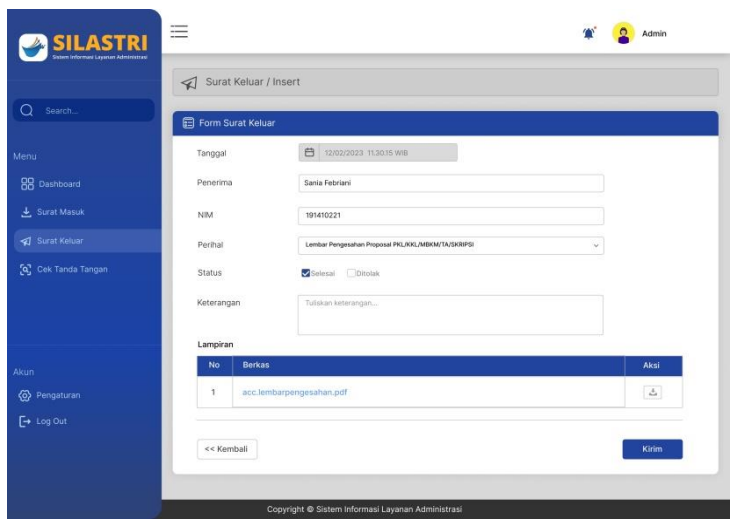

**Gambar 8.** Penerapan Prinsip *Law of Closure*

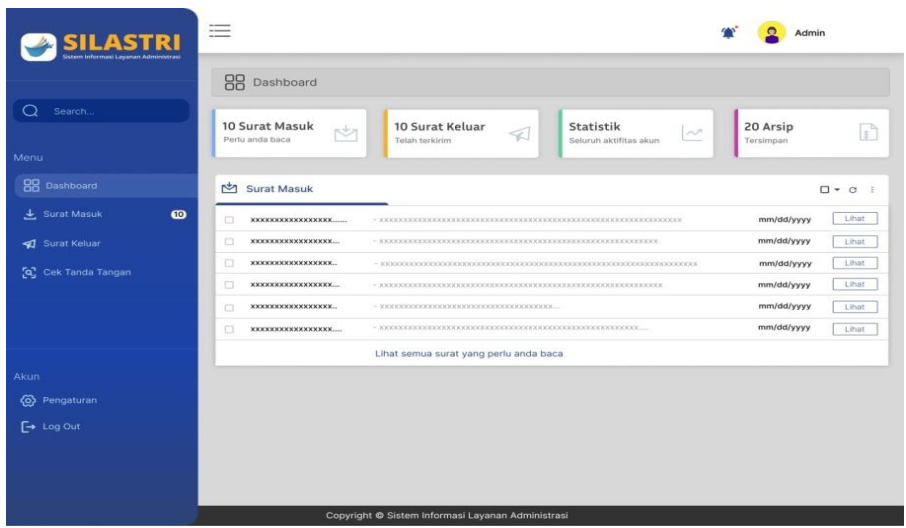

**Gambar 9.** Penerapan Prinsip *Law of Simplicity*

8. *Law of Similarity*

Prinsip *similarity* mirip dengan *proximity*, prinsip ini diterapkan pada semua halaman aplikasi SILASTRI untuk menjaga konsistensi bentuk ataupun kegunaan pada informasi yang diberikan sistem kepada pengguna. Kesamaan tersebut berupa penggunaan tipe font inter di semua layout, tombol dengan satu tipe shape yaitu "rectangle" dan penggunaan warna dominan yaitu biru yang dikombinasikan dengan warna hitam dan putih pada aplikasi.

Sebelum dilakukan pengujian, peneliti membuat tugas dan skenario yang akan dilakukan pengguna. Masing-masing tugas tersebut jika selesai dilaksanakan memiliki catatan waktu penyelesaiannya. Catatan waktu tersebut akan menjadi tolak ukur untuk mengetahui apakah pengguna dapat menggunakan aplikasi tersebut dengan mudah atau tidak. Daftar skenario pengguna dapat dilihat pada tabel berikut.

**Tabel 1.** Skenario *Tugas Usability Testing*

| No | Tugas | Skenario |
|---|---|---|
| 1 | Melakukan *Login* | Anda diharuskan login untuk menggunakan fitur aplikasi. Anda dapat login dengan akun SisFo untuk masuk ke aplikasi. |
| 2 | Melakukan pengajuan dokumen | Anda ingin meminta tanda tangan Ketua Program Studi dan Dekan pada dokumen anda. Gunakan aplikasi untuk melakukan pengajuan. |
| 3 | Melihat status pengajuan | Anda ingin mengecek pengajuan anda apakah sudah disetujui atau belum. Buka aplikasi untuk melihat rincian pengajuan anda. |
| 4 | Mengunduh pengajuan yang di setujui | Dokumen anda sudah disetujui oleh admin, anda ingin mengunduh dokumen tersebut. Gunakan aplikasi untuk mengunduh. |
| 5 | Mengirim kritik dan saran | Anda melihat atau mengalami masalah atau ingin menyampaikan saran. Gunakan aplikasi untuk melakukan menyampaian apa yang anda alami. |
| 6 | Melihat detail pengajuan ditolak | Ketika anda melihat status pengajuan, dokumen anda ditolak dan anda ingin melihat detail alasan dokumen anda ditolak. Gunakan aplikasi untuk melihat detail tersebut. |
| 7 | Melakukan *Logout* | Anda ingin keluar dari aplikasi SILASTRI. Gunakan aplikasi untuk melakukan Logout |

Ketika responden selesai mencoba *prototype*, Maze menampilkan *report* yang terdiri dari *breakdown* skor tiap tugas. Maze otomatis melakukan kalkulasi perhitungan skor, kemudian menghasilkan skor akhir nilai *usability prototype* aplikasi SILASTRI yaitu sebesar 90 dari 30 responden yang menyelesaikan 7 tugas. Setelah responden mencoba prototype SILASTRI, mereka mengisi kuesioner yang terdiri dari 10 pernyataan (Brooke, 1995).

**Tabel 2.** Pernyataan SUS Menurut John Brooke (1995)

| No | Komponen |
|---|---|
| 1 | Saya berpikir akan menggunakan sistem ini lagi. |
| 2 | Saya merasa sistem ini rumit untuk digunakan. |
| 3 | Saya merasa sistem ini mudah digunakan. |
| 4 | Saya berpikir saya butuh bantuan teknisi dalam menggunakan sistem ini. |
| 5 | Saya merasa fitur-fitur sistem ini terintegrasi dengan baik. |
| 6 | Saya merasa ada banyak hal yang tidak konsisten. |
| 7 | Saya merasa orang lain akan memahami cara menggunakan sistem ini dengan cepat. |
| 8 | Saya merasa sistem ini rumit untuk digunakan. |
| 9 | Saya merasa percaya diri dalam menggunakan sistem ini. |
| 10 | Saya perlu membiasakan diri terlebih dahulu sebelum menggunakan sistem ini. |

Dari tiap pilihan responden memiliki skala nilai 5 hingga 1, peneliti menghitung jumlah nilai dari tiap pernyataan. Perhitungan nomor ganjil diperoleh nilai pernyataan responden dikurangi 1, dan untuk pernyataan bernomor genap 5 dikurangi nilai pernyataan responden, penulis menggunakan Microsoft Excel untuk melakukan perhitungan ini. Jumlah tersebut kemudian dikalikan dengan 2,5 sehingga diketahui total skor SUS dengan rumus berikut:

$$\bar{x} = \frac{\sum x}{n}$$

Keterangan :
x    = nilai rata-rata
∑    = jumlah nilai SUS
n    = jumlah responden

**Tabel 3.** Nilai Akhir SUS

| Responden | Nilai Raw Penyataan Ganjil | Nilai Raw Pernyataan Genap |
|---|---|---|
| **70 Responden** | 332 | 96 |
|  | 331 | 110 |
|  | 330 | 83 |
|  | 329 | 98 |
|  | 324 | 183 |
| **Jumlah Nilai Raw** | 1646 | 571 |
| **Jumlah Nilai Keseluruhan** | 2217 | |
| **Total Skor SUS** | 6188 | |
| **Skor Rata-Rata (Hasil Akhir)** | 88 | |

Setelah diketahui total skor SUS yaitu 6188, kemudian dibagi dengan jumlah responden yaitu 70. Nilai skor rata-rata dari prototipe aplikasi SILASTRI adalah 88. Ketika di interpretasikan dalam skala nilai SUS, *prototipe* aplikasi SILASTRI bernilai B, yang masuk kategori *excellent* dan a*cceptable.*

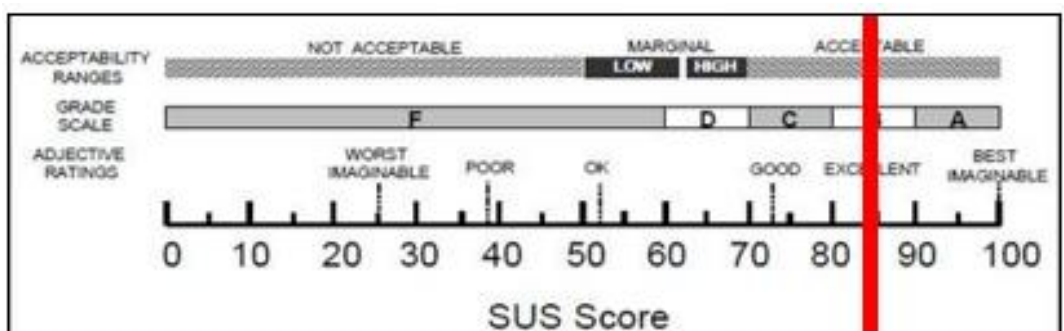

**Gambar 8.** Skor Kuesioner SUS SILASTRI

## KESIMPULAN DAN REKOMENDASI

Pelayanan administrasi mahasiswa Program Studi Sistem Informasi masih bersifat konvensional. Dosen Program Studi tidak memiliki bukti otentik salinan arsip dari surat masuk dan surat keluar sehingga sulit dilakukan *monitoring* sedangkan mahasiswa yang memiliki arsip tersebut hanya berupa lembaran asli yang beresiko rusak atau hilang. Hasil pengujian *prototype* melalui tools Maze memperoleh skor sebesar 89, selain itu melalui perhitungan kuesioner *System Usability Scale* (SUS) mendapatkan *adjective rating* sebesar 88 dengan kategori *good, grade scale* B dan *acceptable*. Oleh karena itu dapat disimpulkan rancangan UI/UX aplikasi SILASTRI dengan menerapkan perspektif psikologi memiliki *interface* dan *user experience* yang diterima dengan baik oleh pengguna.

Hasil dari perancangan ini dapat dijadikan acuan dalam pengembangan aplikasi Sistem Informasi Layanan Administrasi (SILASTRI) dengan implementasi *coding* sehingga dapat langsung digunakan mahasiswa Universitas Bina Darma. Agar aplikasi Sistem Informasi Layanan Administrasi (SILASTRI) lebih berkembang diperlukan penambahan berbagai fitur-fitur yang dapat memudahkan seluruh mahasiswa di Universitas Bina Darma Palembang. Peneliti juga menyarankan untuk menambah jumlah responden agar didapatkan lebih banyak ulasan dan penilaian terhadap rancangan aplikasi yang dibuat.

## REFERENSI